\documentclass[conference]{IEEEtran}
\IEEEoverridecommandlockouts

\pdfoutput=1
\usepackage{graphicx}
\usepackage{cite}
\usepackage{amsmath,amssymb,amsfonts}
\usepackage{pifont}
\usepackage{algorithmic}
\usepackage{textcomp}
\usepackage{multirow}
\usepackage{threeparttable}
\usepackage{lscape}
\usepackage[table,xcdraw]{xcolor}
\usepackage{graphicx}
\usepackage{soul}
\usepackage[textsize=scriptsize,backgroundcolor=yellow!40]{todonotes}
\usepackage{fontawesome}
\usepackage{tikz}
\usepackage{ifthen}
\usepackage{hyphenat}
\usepackage{booktabs}
\usepackage{caption}
\usepackage{enumitem}
\usepackage{subcaption}
\usepackage[multiple]{footmisc}
\usepackage{booktabs}
\usepackage{url}
\usepackage[normalem]{ulem}
\useunder{\uline}{\ul}{}

\usepackage[intoc, spanish]{nomencl}
\makenomenclature

\usepackage{adjustbox}

\newcommand{\commentblock}[1]{}

\def\BibTeX{{\rm B\kern-.05em{\sc i\kern-.025em b}\kern-.08em
    T\kern-.1667em\lower.7ex\hbox{E}\kern-.125emX}}
\begin{document}

\title{Analyzing the Performance of the Inter-Blockchain Communication Protocol
}

\author{\IEEEauthorblockN{João Otávio Chervinski$\mathsection$\dag, Diego Kreutz$\mathsection\ddagger$, Xiwei Xu\dag, Jiangshan Yu$\mathsection$}
\IEEEauthorblockA{$\mathsection$\textit{ Monash University, Australia}\\ \dag\textit{ CSIRO's Data61, Australia}} $\ddagger$ \textit{Federal University of Pampa, Brazil}}

\maketitle

\begin{abstract}

With the increasing demand for communication between blockchains, improving the performance of cross-chain communication protocols becomes an emerging challenge. We take a first step towards   analyzing the limitations of cross-chain communication protocols by comprehensively evaluating Cosmos Network's Inter-Blockchain Communication Protocol. To achieve our goal we introduce a novel framework to guide empirical evaluations of cross-chain communication protocols. We implement an instance of our framework as a tool to evaluate the IBC protocol. Our findings highlight several challenges, such as high transaction confirmation latency, bottlenecks in the blockchain's RPC implementation and concurrency issues that hinder the scalability of the cross-chain message relayer. We also demonstrate how to reduce the time required to complete cross-chain transfers by up to 70\% when submitting large amounts of transfers. Finally, we discuss  challenges faced during deployment with the objective of contributing to the development and advancement of cross-chain communication.

\end{abstract}


\section{Introduction} 

Blockchains were designed to enable decentralized digital currencies, but evolved into technology for avoiding censorship, managing decentralized organizations and providing data transparency. The global blockchain market was valuated at USD 5.92 billion in 2021 and is predicted to grow as blockchain applications continue to bring value to a wide range of industries~\cite{grandviewblockchain}.

While the interest in distributed ledger technology has increased, long-standing blockchain issues such as low throughput, high latency and trade-offs between security, scalability and decentralization~\cite{vukolic2015quest, croman2016scaling, gervais2016security} are still prevalent. For instance, Bitcoin and Ethereum achieve a transaction throughput of roughly 7 and 15 transactions per second (tps) respectively, and are often compared to traditional payment systems such as Visa, which claims to be capable of processing 24,000 tps\footnote{https://usa.visa.com/run-your-business/small-business-tools/retail.html}, to highlight the large gap between their performance. 

Several works have evaluated and identified issues affecting the performance of both public and permissioned blockchains~\cite{aldweesh2019opbench,baliga2018fabric,baliga2018quorum,dabbagh2020performance,dinh2017blockbench,hyperledgercaliper,wang2019performance,fan2020performance,kuzlu2019performance,saingre2020bctmark,xu2021latency,wang2021xbcbench,thakkar2018performance,sedlmeir2021dlps, gramolidiablo2023}. For instance, Hyperledger Fabric v0.6.0 is unable to scale beyond 16 nodes because of implementation issues in its consensus protocol~\cite{dinh2017blockbench}. 

Similarly, previous work shows that certain cryptography operations and REST API calls have led to transaction validation bottlenecks in Hyperledger Fabric v1.0~\cite{thakkar2018performance}. Beyond evaluating the system's performance, the authors propose optimizations that lead to a 16x increase in throughput. This highlights the importance of in-depth performance evaluations.

In modern blockchain applications, which increasingly require  exchanging data between different blockchains (a.k.a. cross-chain communication)~\cite{forbes2022interoperability, cointelegraph2022interoperability,abebe2019enabling, gordon2018blockchain}, bottlenecks may have a severe impact as they degrade the performance and increase waiting times for all blockchains participating in an operation. Cross-chain communication requires specific protocols to connect and reliably deliver data between blockchains, however, those protocols introduce additional points of failure and new bottlenecks, more so given that many are still under active development~\cite{belchior2020survey, johnson2019sidechains}.

Various protocols have been proposed to enable cross-chain communication~\cite{herlihy2018atomic,kiayias2019proof,gazi2019proof, ethrelay, karantias2019proof, kwon2019cosmos, wood2016polkadot, nick2020liquid, interledgerprotocol, btcrelay, zamyatin2019xclaim, back2014enabling, abebe2020verifiable}, each with different trade-offs between security, performance and deployment complexity~\cite{chervinski2022characterizing}. Relay protocols such as BTC Relay~\cite{btcrelay} and ETH Relay~\cite{ethrelay} employ smart contracts to verify transactions added to external blockchains, but work only one-way and are costly to maintain. Sidechains~\cite{back2014enabling, kiayias2019proof, gazi2019proof, nick2020liquid} enable two blockchains to send assets back and forth via a two-way peg mechanism, but must be designed for a specific pair of blockchains. Blockchain ecosystems, such as Cosmos~\cite{kwon2019cosmos} and Polkadot~\cite{wood2016polkadot}, host independent blockchain applications and provide protocols to facilitate cross-chain communication, but impose constraints on blockchains to make them compatible with the ecosystem.

Cross-chain operations differ from those executed in isolated blockchains in the following ways: they must undergo consensus to be recorded in all participating blockchains; they require a reliable communication channel through which data can be sent from one blockchain to another; they might be rejected by a participating blockchain, therefore they require atomicity in order to maintain a consistent state across ledgers~\cite{han2019optionality, hardjono2021blockchain, belchior2022hermes}. Hence, cross-chain performance analysis requires an approach that is different from those proposed by existing evaluation frameworks for isolated blockchains.

Among proposals to enable cross-chain communication, Cosmos (3 billion USD market cap) and Polkadot (6.2 billion USD market cap)\footnote{https://coinmarketcap.com/} have attracted significant attention in the past few years. Cosmos is a network of blockchains and decentralized applications built using its own SDK. Blockchains deployed in the Cosmos network can leverage a hub-and-spoke topology and the Inter-Blockchain Communication (IBC) protocol to exchange information. Binance Chain, Osmosis, Secret Network and Ethermint\footnote{https://v1.cosmos.network/ecosystem/apps} are examples of chains deployed in the Cosmos ecosystem. Similarly, Polkadot is a network of interconnected blockchains, called parachains, developed with the Substrate framework. Parachains can communicate through the XCMP (Cross-Chain Message Passing) protocol, built on top of the XCM (Cross-consensus Message) format. The Moonbeam network is currently deployed in the Polkadot ecosystem and a Chainlink implementation is currently under development\footnote{https://parachains.info/}. But, despite their popularity, there are no performance evaluations of cross-chain communication using IBC or XCMP. This may be partly explained by fact that those protocols are under active development, which leads to breaking changes, outdated documentation and scarcity of information regarding setup and configuration.

In this paper, we present the first performance analysis of the Inter-Blockchain Communication protocol paired with the Hermes Relayer, both of which are currently used to enable cross-chain communication in the Cosmos Network. IBC is an open source protocol designed to facilitate communication between heterogeneous blockchains. It defines a standard for constructing cross-chain messages and requires only the implementation of a minimal set of functions in the communicating blockchains. This gives IBC the potential to be used not only between blockchains deployed in the Cosmos ecosystem, but any blockchains that meet the requirements. Unlike XCMP, which is currently under active development and being replaced by a temporary protocol~\cite{polkadotlearnxcmp}, working implementations of IBC are already being used to connect over 50 homogeneous blockchains including the Cosmos Hub, Osmosis and the Juno Network\footnote{https://hub.mintscan.io/chains/ibc-network}. Additionally, Polymer Labs and Hyperledger YUI Labs are working toward supporting IBC on existing blockchain platforms such as Ethereum, Hyperledger Fabric, Hyperledger Besu, Corda, Solana, Polygon and Fantom~\cite{yuilabs, polymerlabsibc}.

Evaluating the performance of cross-chain communication protocols such as IBC poses additional challenges compared to isolated blockchains: 
\begin{enumerate}[label=(\roman*)]
\item The protocol is composed of multiple steps (\textit{transfer, receive, acknowledge and timeout}) that must be recorded in the two communicating blockchains. 

\item Blockchains are unable to communicate directly and require additional agents and/or protocols to deliver messages from one blockchain to another. 

\item Operations may fail after having steps recorded in the blockchain. For instance, when a message is not delivered before timing out, intermediary steps recorded in transactions remain in the blockchain even though the cross-chain operation was not completed. 
\end{enumerate}

We emphasize that (i) increases the complexity of the experimental setup as multiple systems need to be deployed and connected to each other. Similarly, (ii) and (iii) increase analysis complexity by requiring the state of operations to be tracked across all of the systems involved in cross-chain communication.

In this paper we make four major contributions:

\begin{itemize}
    \item We analyze the performance of cross-chain communications between Cosmos Gaia blockchains using the Inter-Blockchain Communication protocol and the Hermes Relayer\footnote{https://github.com/informalsystems/ibc-rs}. Throughout our comprehensive throughput, latency and relayer scalability analysis we identify several surprising results and issues impairing performance. First, we show that using two relayers to relay for a single cross-chain channel decreases throughput by 33\% compared to using a single relayer. Second, we show that the relayer application processes cross-chain transfers in batches, leading to high transfer completion latency, e.g., 455 seconds for 5,000 transfers. Third, we show that the main cross-chain communication bottleneck, which lies in the blockchain's RPC implementation, causes 69\% of the time required to process cross-chain transfers to be spent on querying data from the blockchains.
    
    \item  We propose a novel framework for evaluating the performance of cross-chain communication protocols. As a first instantiation of our framework, we implement and make available an open source tool~\cite{code-data} to measure the performance of cross-chain communication between Cosmos blockchains using the IBC protocol. Our tool reduces the effort required to deploy and connect two Cosmos blockchains using the Hermes Relayer and provides seven configurable parameters that can be used to evaluate different blockchain configurations. Additionally, our tool generates execution reports to assist in performance evaluations for different setup configurations.
    
    \item We discuss five challenges faced during the deployment of the Cosmos Gaia blockchains and the Hermes Relayer. In Section~\ref{sec:challenges} we discuss how those challenges can impact cross-chain performance and increase the difficulty of using such kind of systems.

    \item We provide a 158GB dataset of execution logs to assist in future research~\cite{code-data}.

\end{itemize}

\section{The Cosmos Ecosystem}

Cosmos is a network (also referred to as an ecosystem) of independent application-specific blockchains, called zones, interconnected through blockchains named hubs. Cosmos zones are powered by Tendermint BFT, a deterministic, partially synchronous consensus engine~\cite{buchman2018latest}. Hubs represent points of connection between zones, allowing the transfer of information. Different zones can communicate via the Inter-Blockchain Communication (IBC) protocol~\cite{goes2020interblockchain}.

Cosmos provides a modular open-source framework, the Cosmos SDK, for the development of both permissioned and permissionless blockchains and applications. 
The Cosmos SDK enables developers to build on top of the Tendermint platform and aims to shorten development time by providing access to open-source composable modules that implement common blockchain functionalities such as token minting, token transfers, staking and punishment for misbehaviour.

\subsection{Tendermint}

Tendermint is a blockchain platform that provides a consensus engine called Tendermint Core and an interface called Application BlockChain Interface (ABCI). Tendermint Core consists of a Byzantine Fault Tolerant, deterministic and partially synchronous consensus engine. It also provides a peer-to-peer networking protocol which enables blockchain nodes to communicate. Tendermint's ABCI provides a generic interface that allows blockchain applications to interact with the consensus engine provided by Tendermint Core.

Tendermint's consensus engine provides a protocol through which network participants called validators can cooperate to append information to the blockchain and update its state. The protocol can tolerate arbitrary behavior from up to one third of the network's validators while still guaranteeing consistency of state among honest nodes. 

The protocol to achieve consensus is executed in rounds.
In each round one participant from the validator set is elected as a proposer and gets to suggest a block of transactions for the current blockchain height. The remaining validators take part in two stages of voting, named pre-voting and pre-commiting, to decide whether to accept the proposed block or not. 
As soon as 2/3 of the validators cast their pre-vote for the same block, the protocol can receive pre-commits.
If a block receives pre-commits from 2/3 of the validators before the end of the round, it is appended to the blockchain and the state of the ledger is updated based on its contents. If timeout is reached during a round and validators do not agree on a new block, the protocol advances to the next round and a new validator has the chance to propose a block. 

\begin{figure}[!htb]
  \centering
  \includegraphics[width=0.99\columnwidth]{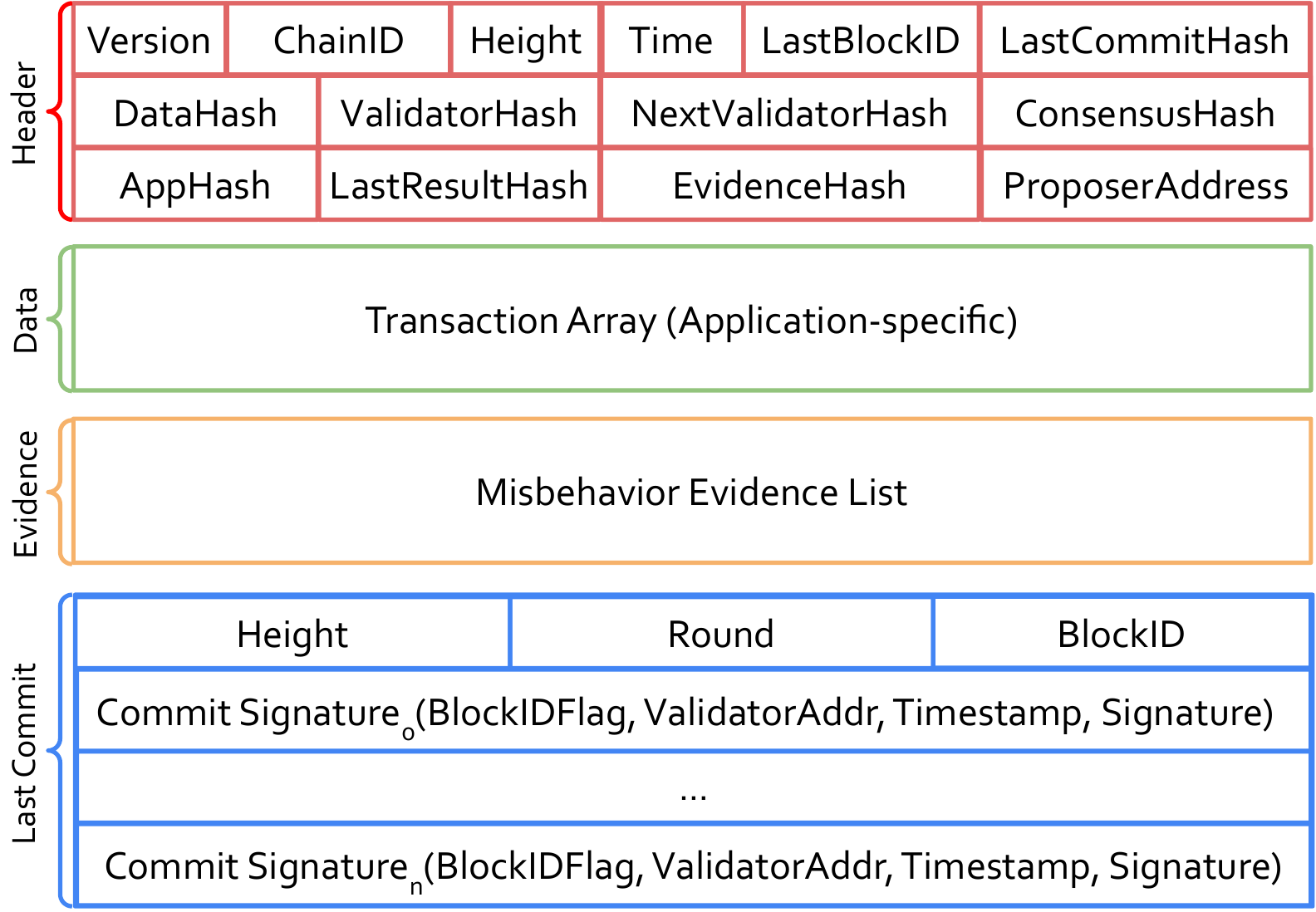}
  \caption{Structure of a Tendermint block.}
  \label{fig:tendermint_block}
\end{figure}

We show the structure of a Tendermint block in Figure \ref{fig:tendermint_block}. A block is divided into four main fields, the \textit{Header}, the \textit{Data} field, the \textit{Evidence} field and the \textit{LastCommit} field. Except for the \textit{Data} field, Tendermint performs validation on all the fields that compose a block. Transaction data is application-specific and unknown to Tendermint, hence it is a responsibility of the blockchain application to perform the validation of this data.

The \textit{Header} contains information regarding the block and its position on the chain, as well as metadata related to consensus, validator set and the blockchain application. The \textit{Data} field contains the set of transactions chosen by the block proposer and agreed upon by the consensus validators. Proofs of malicious activity carried out by members of the validator set are included in the \textit{Evidence} field. Those proofs can be used by the blockchain application to punish actors who deviate from the protocol. This field is empty in the absence of misbehavior.

Lastly, the \textit{LastCommit} field contains the votes from the participants of the validator set. It contains data regarding block height, consensus round information, a \textit{BlockID} of the corresponding block and an array of validator signatures. The signatures include the \textit{BlockIDFlag}, indicating whether the validator voted for the block accepted by the majority of validators, \textit{nil}, for a different block or did not cast a vote. Together with a vote, a validator's address, timestamp and commitment signature are included.

\subsection{IBC Protocol}

The Inter-Blockchain Communication protocol is an open-source, end-to-end, payload agnostic protocol that provides reliable and authenticated message passing between IBC modules residing in distinct blockchains~\cite{goes2020interblockchain}. Aside from a set of functions described in the interchain standard that describes host requirements (ICS 24)\footnote{https://github.com/cosmos/ibc/tree/master/spec/core/ics-024-host-requirements}, no other prerequisites are imposed on the host blockchains.

Data transported across blockchain modules is opaque to the IBC protocol and only needs to be readable by the sending and receiving applications. This makes the protocol flexible and enables it to be applied for a variety of use cases such as fungible token transfers, deployment of multi-chain contracts, multi-chain account management and cross-chain data sharing.

The IBC protocol handles authentication, ordering and transport of data, but it does not have access to a transport layer to deliver data to other blockchains. Hence, in order to communicate using IBC, blockchains require access to external applications, called relayers, with access to network infrastructure and communication protocols such as TCP/IP.

Relayers are off-chain processes responsible for scanning blockchain data to identify messages that need to be delivered to another blockchain.
When pending messages are found, the relayer retrieves them from the blockchain, transforms them into IBC datagrams and delivers them to their destination.

\subsubsection{IBC Channel Setup}

Blockchains can only communicate via IBC once a channel is established between them. Channels function as routes between communicating blockchain modules and, as per the IBC protocol's definitions, are responsible for providing ordering, exactly-once delivery and controlling the permissions that determine which modules are able to send and receive packets.

To open an IBC channel, a \textit{connection} between the pair of communicating blockchains is required. Connections are established through a handshaking process and require both blockchains to run a light client of the counterparty chain. Light clients keep track of the consensus information of the blockchain they are monitoring and enable the verification of state updates. 
Once the light clients are running, the blockchains can open a \textit{connection} enabling access to authorised communication.

After going through the handshaking process and establishing a \textit{connection}, a pair of communicating blockchains can open a \textit{channel} through which they can send and receive IBC packets. Channels can be either ordered, delivering packets in the order they are sent, or unordered, where packets are delivered in any order. Two blockchains can open multiple \textit{channels} using a single \textit{connection}. 

\subsubsection{The IBC packet life cycle}

IBC messages must be included in the blockchain before being sent through a cross-chain channel. The relayer application scans new blocks appended to the blockchain looking for transactions with IBC messages addressed to the channels it relays for. Upon finding pending messages, the relayer extracts their information from the blockchain, builds IBC packets and delivers them to the destination blockchain. A cross-chain transfer requires the inclusion of 3 messages in the communicating blockchains to be completed, namely, the transfer message (\textit{MsgTransfer}), the receive message (\textit{MsgRecvPacket}) and the acknowledgement message (\textit{MsgAcknowledgement}). The transfer message requests a fungible token transfer and stores a proof of commitment to the packet data and packet timeout in the sending blockchain. Upon receiving a packet with a transfer request, the blockchain to which it is addressed verifies the packet commitment proof in the state of the sending chain and includes a transaction containing a receive message to accept it. Based on the receive message, an acknowledgement is sent back to the blockchain that initiated the transfer. Upon receiving the acknowledgement, the blockchain verifies that a proof of acknowledgement was stored in the receiving chain and includes a transaction with an acknowledgement message. It is worth emphasizing that relayers must be connected to full nodes of both blockchains to execute queries for transaction data and relay packets.

\begin{figure}[!htb]
  \center
  \hspace*{.3cm}
  \includegraphics[scale=0.38]{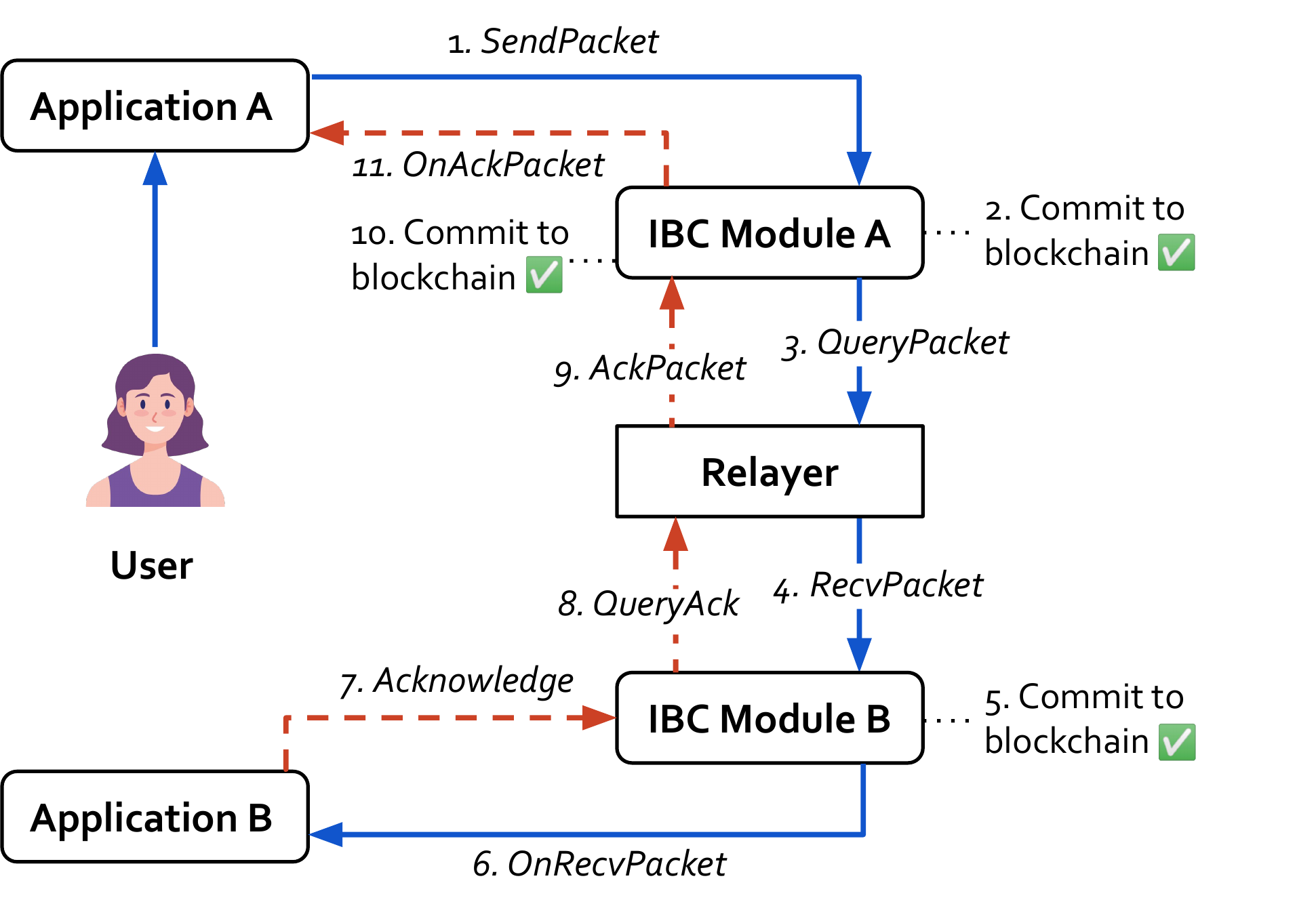}
  \caption{IBC protocol packet flow (sucessful).}
  \label{fig:ibc_packet_flow}
\end{figure}

We show the flow of a packet sent via the IBC protocol in Figure \ref{fig:ibc_packet_flow}. Let blockchain$_{A}$ and blockchain$_{B}$ be a pair of blockchains. Blockchain$_{A}$ runs application$_{A}$ and implements a module that handles inter-blockchain communication, called IBC module$_{A}$. Similarly, blockchain$_{B}$ runs application$_{B}$  and implements IBC module$_{B}$.

Firstly, a user requests application$_{A}$ to send a packet to application$_{B}$. Application$_{A}$ then calls the \textit{SendPacket} function implemented by IBC module$_{A}$ on the same blockchain. IBC module$_{A}$ stores a commitment to the outgoing packet and its timeout. This commitment is added to the state of the blockchain. 
Secondly, the relayer application queries the blockchain for pending messages and retrieves the information it needs to construct the corresponding IBC packet. Once the relayer has a packet ready for delivery, the receiving module on the destination blockchain, IBC module$_{B}$, triggers the \textit{RecvPacket} function. 
Thirdly, the receiving module verifies the packet's timeout and the commitment in the source chain. If the data is successfully validated, the module routes the packet to application$_{B}$. Fourthly, application$_{B}$ processes the operations contained in the packet's data and sends an acknowledgement back to IBC module$_{B}$. Fifthly, this module stores a commitment acknowledging that application$_{B}$ has received the packet from application$_{A}$. This commitment is stored in the state of the blockchain and is then queried by the relayer application. Next, the relayer builds an \textit{AcknowledgePacket} and delivers it to IBC module$_{A}$ in the source blockchain. At this point, packet commitments stored by the module can be deleted since the packet has been successfully delivered.

\begin{figure}[!htb]
  \centering
  \includegraphics[scale=0.38]{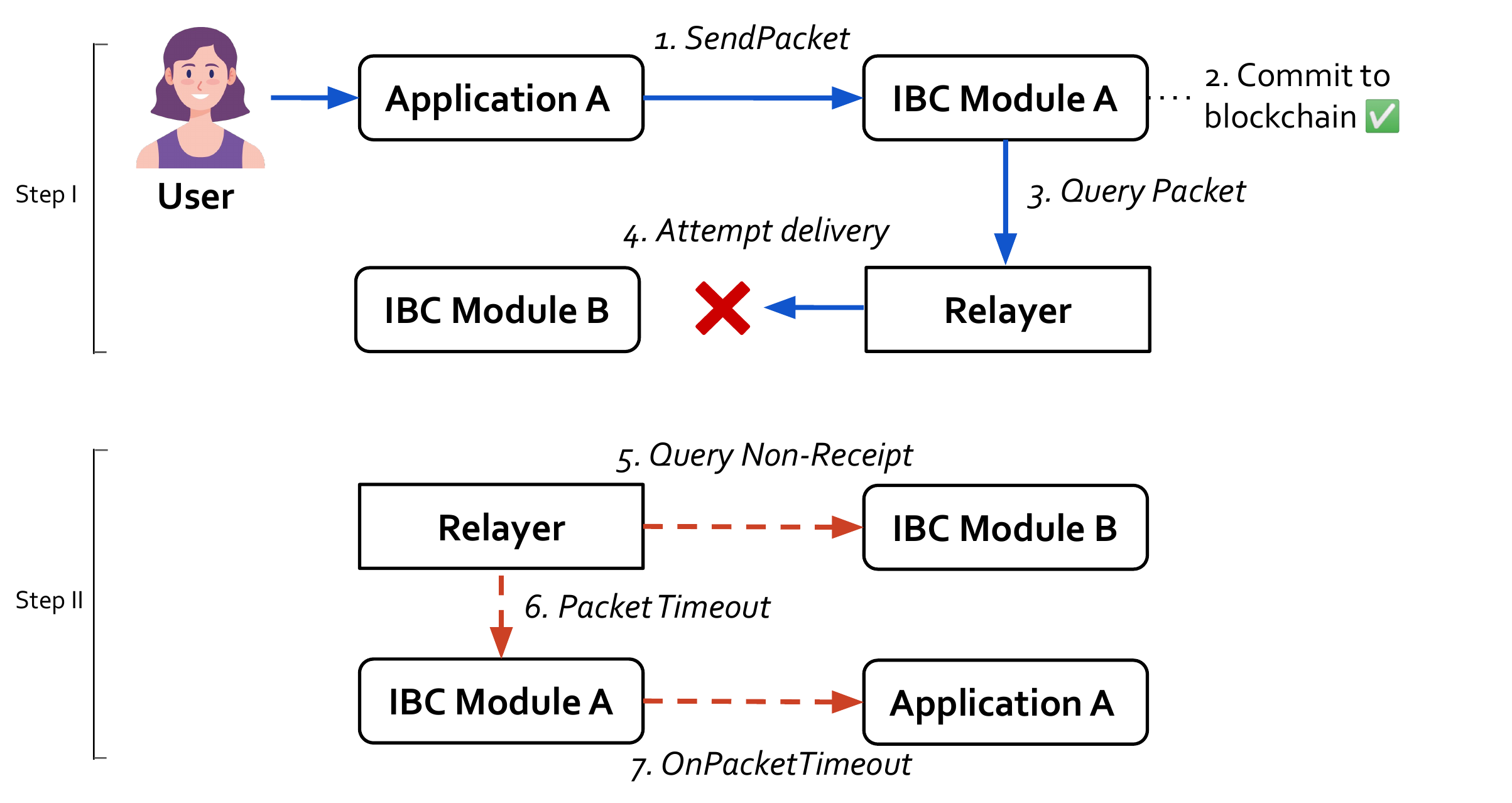}
  \caption{IBC Protocol packet flow (timeout).}
  \label{fig:ibc_packet_flow_timeout}
\end{figure}

Due to the nature of network asynchrony, packets may not be delivered before timing out. Timeouts can also be triggered when the receiving blockchain module does not accept the incoming packet. We illustrate the IBC packet with expired timeout in Figure \ref{fig:ibc_packet_flow_timeout}. 
Application$_{A}$ calls the \textit{SendPacket} function and IBC module$_{A}$ stores the packet commitment and timeout. However, it fails to be delivered to the destination module on blockchain$_{B}$ before it reaches the specified timeout. The relayer application identifies that the packet can no longer be accepted by the destination and sends a query to IBC module$_{B}$ asking for a proof that it has not been received. This information is sent back to IBC module$_{A}$ in the source blockchain and used to trigger the \textit{OnPacketTimeout} function. This function implements the logic for undoing operations executed before packet commitment, such as unlocking assets that were previously held locked while the transfer request was pending.

\subsection{Hermes Relayer}

The Hermes Relayer\footnote{https://github.com/informalsystems/hermes} is an open-source implementation of an IBC relayer written using the Rust programming language. It is one of the two IBC relayers that are currently under active development, the other being the Golang Relayer\footnote{https://github.com/cosmos/ibc-go}.
We chose to use the Hermes Relayer in our analysis of cross-chain communication for three reasons. First, it offers comprehensive documentation. Second, it provides extensive event logging capabilities, a feature that is essential for performance analysis. Lastly, at the time of writing, this relayer offers more features and is updated more often than the Golang Relayer. 

\begin{figure}[!htb]
  \centering
  \includegraphics[width=0.98\linewidth]{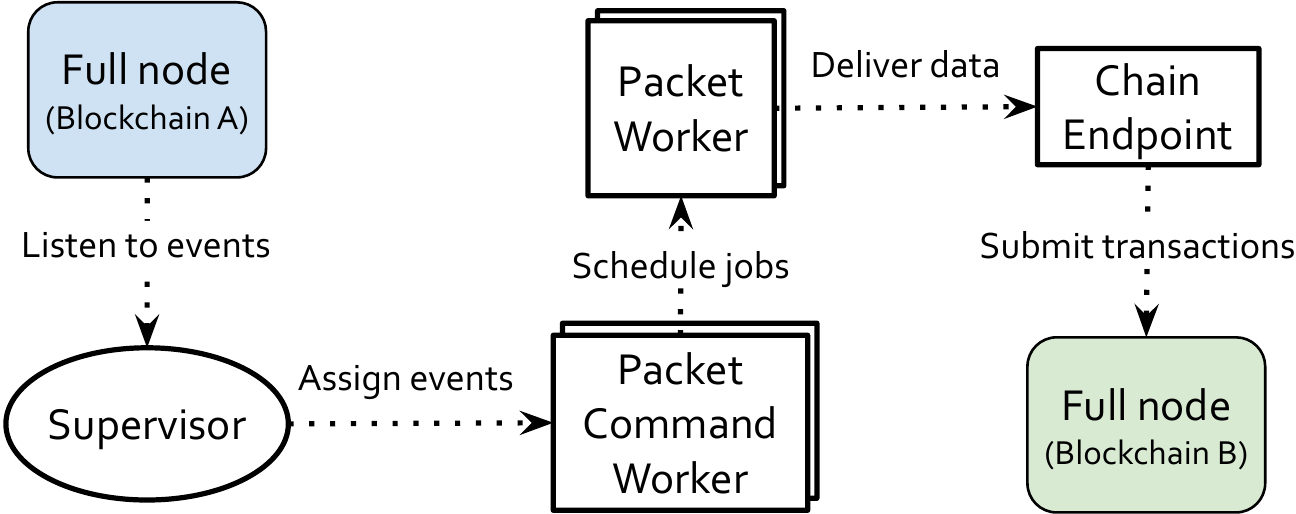}
  \caption{Architectural overview of the Hermes Relayer.}
  \label{fig:hermes_architecture}
\end{figure}

In Figure \ref{fig:hermes_architecture}
we present an overview of the  Hermes Relayer's architecture. The Supervisor subscribes to events generated by blockchain full nodes. Each event is forwarded to the Packet Command Worker associated with its cross-chain channel. The Packet Command Worker schedules relayer tasks and assigns them to Packet Workers. Upon completing the assigned tasks, Packet Workers forward the resulting data to the Chain Endpoint, which interfaces directly with the destination blockchain through a full node. We refer the reader to the developer's overview~\cite{hermesv1architecture} for an in-depth discussion of the Hermes Relayer's architecture.

\section{Experimental Methodology}\label{sec:experimental_methodology}

We first introduce a novel framework for evaluating cross-chain communication protocols. 
We provide a first instantiation of our framework as a tool for evaluating the performance of the IBC protocol when used for cross-chain communication between two Cosmos Gaia blockchains.
Our tool provides assistance in experiment setup, execution, data collection and data analysis. For our experiments we set up a private testnet environment and use it to measure the performance of the IBC protocol under different scenarios.

\subsection{Cross-chain Performance Evaluation Framework}

\begin{figure}[!htb]
  \centering
  \includegraphics[width=0.98\linewidth]{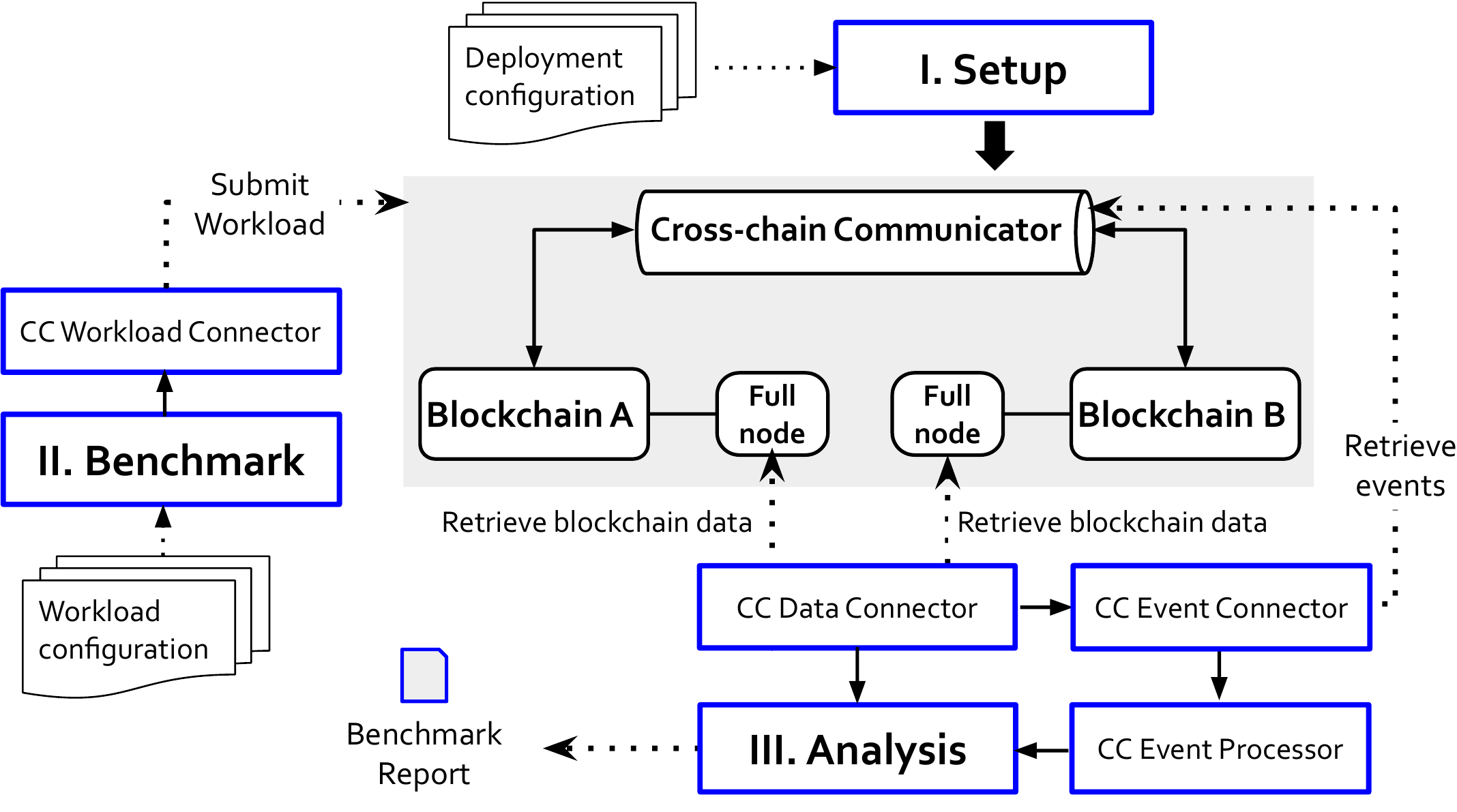}
  \caption{Cross-chain performance evaluation framework.}
  \label{fig:cc_framework}
\end{figure}

We present our framework for evaluating the performance of cross-chain communication protocols in Figure \ref{fig:cc_framework}. To the best of our knowledge, this is the first framework to guide empirical evaluation of cross-chain communication performance.

Our framework introduces four new key components, namely the \textit{ Cross-chain Communicator}, \textit{Cross-chain Data Connector}, \textit{Cross-chain Event Connector} and the \textit{Cross-chain Event Processor}. The \textit{Cross-chain Communicator} enables data transfer between blockchains; the \textit{Cross-chain Data Connector} is required to retrieve data from different communicating blockchains concurrently; the \textit{Cross-chain Event Connector} monitors and collects events associated with cross-chain communication, both from the communicating blockchains and from the \textit{Cross-chain Communicator} component; and the \textit{Cross-chain Event Processor} aggregates and interprets cross-chain communication events and their corresponding steps.

Our framework has three main modules, namely \textit{Setup}, \textit{Benchmark} and \textit{Analysis}. The \textit{Setup} module provides flexibility and automation for deploying blockchains with user-chosen parameters. This module is also responsible for configuring and deploying the \textit{Cross-chain Communicator} (e.g, relayers for IBC). If evaluating cross-chain communications in an already deployed environment, the \textit{Setup} can be skipped. 

The \textit{Benchmark} module allows users to evaluate cross-chain communication by executing workloads of different cross-chain operations (e.g, fungible and non-fungible token transfers and cross-chain data queries). It provides the \textit{Cross-chain Workload Connector} component, which uses different resources (e.g., APIs, CLIs) to submit cross-chain operations directly to a blockchain or to an external application (e.g, the Hermes Relayer) that submits it to a blockchain. 

The \textit{Analysis} module processes blockchain and cross-chain communication data to generate performance metrics. This module provides the \textit{Cross-chain Data Connector}, \textit{Cross-chain Event Connector} and \textit{Cross-chain Event Processor}. While in cross-chain communication protocols such as IBC a relayer fulfils the role of the \textit{Cross-chain Communicator}, in Atomic Cross-chain Swaps this role is played by the users participating in the protocol. In the second case the \textit{Cross-chain Event Connector} will retrieve events from the blockchain (e.g, transactions to claim assets from a contract) rather than from the relayer, using the \textit{Cross-chain Data Connector} component.

\subsection{Performance Analysis Tool}\label{sec:tool_design}

As a first instantiation of our framework, we implement a performance analysis tool to evaluate the IBC protocol. Our tool assists with the configuration and automated deployment of the test environment, workload execution and generation of performance metrics.

We implement our \textit{Setup} module to deploy two Cosmos Gaia blockchains and connect them via a cross-chain channel established using the Hermes Relayer. The \textit{Benchmark} module uses cross-chain fungible token transfers and instantiates a \textit{Cross-chain Workload Connector} that binds the workload submission to the Hermes Relayer CLI. For the \textit{Analysis} module, we implement an instance of the \textit{Cross-chain Data Connector} which provides an interface to Tendermint RPC endpoints served by full nodes, allowing us to query data from the communicating blockchains.
We further implement instances of the \textit{Cross-chain Event Connector} and the \textit{Cross-chain Event Processor} to collect and parse logs generated by the Hermes Relayer.

\subsection{Deployment Configuration} 

We set up a private testnet environment composed of two Cosmos Gaia v7.0.3 blockchains, each maintained by five validator nodes. As we show later in Section \ref{sec:performance_evaluation}, the number of consensus nodes has a negligible impact in our analysis as the performance bottlenecks lie in the relayer application and the blockchain's RPC server implementation. The Gaia blockchain natively supports the IBC protocol and was developed to serve as the Cosmos Hub in the Cosmos Network. 

We use Hermes Relayer 1.0.0 to connect the blockchains via an unordered cross-chain channel. The IBC protocol is implemented by application layer modules of the Gaia blockchains and by the Hermes Relayer.
We employ five machines equipped with Intel i7-9700 3GHz processors, 16GB of 2666 MT/s RAM, and 7200RPM HDDs, running the Debian 11 ``bullseye'' operating system and connected to a local area network to run our experiments. We simulate wide area network conditions by enforcing a round-trip latency of 200 milliseconds between any pair of machines, similar to previous works~\cite{yu2019repucoin, kogias2016enhancing, neiheiser2021kauri}. Each machine hosts two full nodes, one validator of the source blockchain and one validator of the destination blockchain. The relayer application is executed in a machine together with two validator nodes, but no more than one instance of the relayer application is executed in each machine. This setup allows the relayer to interact with the blockchain nodes via local endpoints and is recommended for use in production.

While our relayer setup is the same as that of a real application, the number of validators differs, i.e., 5 in our experiments and up to 128 in some blockchains. For a payload of 1024 bytes, consensus latency is approximately 25ms for 5 validators and 110ms for 128 validators~\cite{yin2018hotstuff}. However, we argue that this higher latency has an insignificant impact (approx. 1\%) on the performance of cross-chain communications. For example, completing a single cross-chain transfer (requiring 3 blockchain transactions) takes 21 seconds on average in our experiment. When considering a real application with 128 validators, the added latency for each complete cross-chain transfer is approximately 255ms.

\subsection{Experiment Settings}

We configure the blockchains so that the time interval between the creation of two consecutive blocks is of at least 5 seconds. Blocks containing large amounts of transactions may increase the block interval beyond 5 seconds to allow time for the transactions to be processed. 

We call \textit{source} the blockchain that receives requests to initiate the transfer of tokens (\textit{MsgTransfer}) to another blockchain (referred to as \textit{destination} blockchain). 
Every transaction we submit to the source blockchain contains batches of 100 cross-chain transfer messages. This is the maximum number of messages per transaction allowed by the relayer application. We chose this number of messages for two reasons. First, this facilitates the submission of large amounts of transfers while reducing blockchain transaction processing overhead. Only one transaction has to be validated in order for all the messages within it to be processed. This allows us to include cross-chain transfer requests in the source blockchain at a faster rate, putting  more stress on the relayer application. Second, it helps us in overcoming a limitation\footnote{https://github.com/cosmos/cosmos-sdk/blob/274f389111c323c850c981c\\0de1b7b57eeb23912/x/auth/ante/sigverify.go\#L219-L230} caused by Cosmos blockchains, which restricts the number of transactions each user account (public key) can submit per block to 1. Cosmos' blockchains enforce transaction ordering through the use of account sequence numbers to prevent transaction replay attacks. In its current implementation, this mechanism causes users to have to wait for the confirmation of a submitted transaction before submitting another one with the next sequence number. This prevents us from using multiple transactions from the same user account to increase the number of cross-chain transfer requests within a single block. We use multiple user accounts to submit transactions containing 100 cross-chain transfer messages each to mitigate this problem.

When presenting our experiments, we refer to input rates in terms of requests per second and to throughput in terms of transfers per second. Those input rates correspond to the number of cross-chain transfers we submit per second to the source blockchain in the best case scenario, i.e., when one new block is produced each 5 seconds. If the interval is longer than that, the input rate will be affected. For instance, when the blockchain takes 10 seconds to produce a block, the number of transfers submitted per second is reduced by half. This happens due to the aforementioned limitation on transaction submission, causing the relayer to submit one transaction for each user and wait for its confirmation before submitting the next one. Therefore, a request rate of 1,000 transfers per second corresponds to a batch of 5,000 transfers being submitted every 5 seconds. 
To avoid ambiguity with the widespread abbreviation of transactions per second (TPS), for the remainder of this work we refer to TransFers Per Second as TFPS.

\subsection{Evaluation Metrics}

In this work we use the following performance metrics to evaluate cross-chain communications:

\begin{itemize}
    \item \textbf{Throughput}: measures how many cross-chain transfers are completed per second. A transfer is considered completed only when all three required steps (\textit{transfer, receive, acknowledge}) are correctly recorded in the blockchain.
    
    \item \textbf{Latency}: measures the amount of time required for an operation to be completed. 
    As cross-chain operations usually require several seconds to be completed because of block intervals, we measure latency in seconds.

    \item \textbf{Relayer scalability:} measures the change in throughput and latency taking into account the number of concurrent relayers working for a cross-chain channel.

\end{itemize}
\section{Performance Evaluation}\label{sec:performance_evaluation}

In this section we assess the performance of the IBC protocol using the tool and metrics presented in Section \ref{sec:experimental_methodology}. We analyze 158GB of data generated through 460 hours of experimentation. Based on our findings, we provide observations on how to improve the performance of cross-chain communication. In particular, we show that the main bottleneck lies in the blockchain's RPC server implementation. We also observe that when two relayers work concurrently for the same channel they are unable to relay in a coordinated manner due to their inability to communicate with each other. 
As a consequence, this reduces throughput and hinders scalability in the presence of multiple relayers. Such discoveries are fundamental for the evolution of IBC as well as other cross-chain communication protocols.

\subsection{Throughput}

We analyze the throughput of the different components used to achieve cross-chain communication, namely the Tendermint-based Cosmos Gaia blockchains and the Hermes Relayer. This allows us to identify performance bottlenecks and show how the peak throughput achieved by those components differ.
Specifically, we show that including cross-chain transfer requests in the blockchain can be done at a rate over 10 times faster than the relayer application is able to complete them. This demonstrates that our experiments are stressing the relayer application rather than the blockchains.

\begin{figure}[!htb]
  \centering
  \includegraphics[scale=0.76]{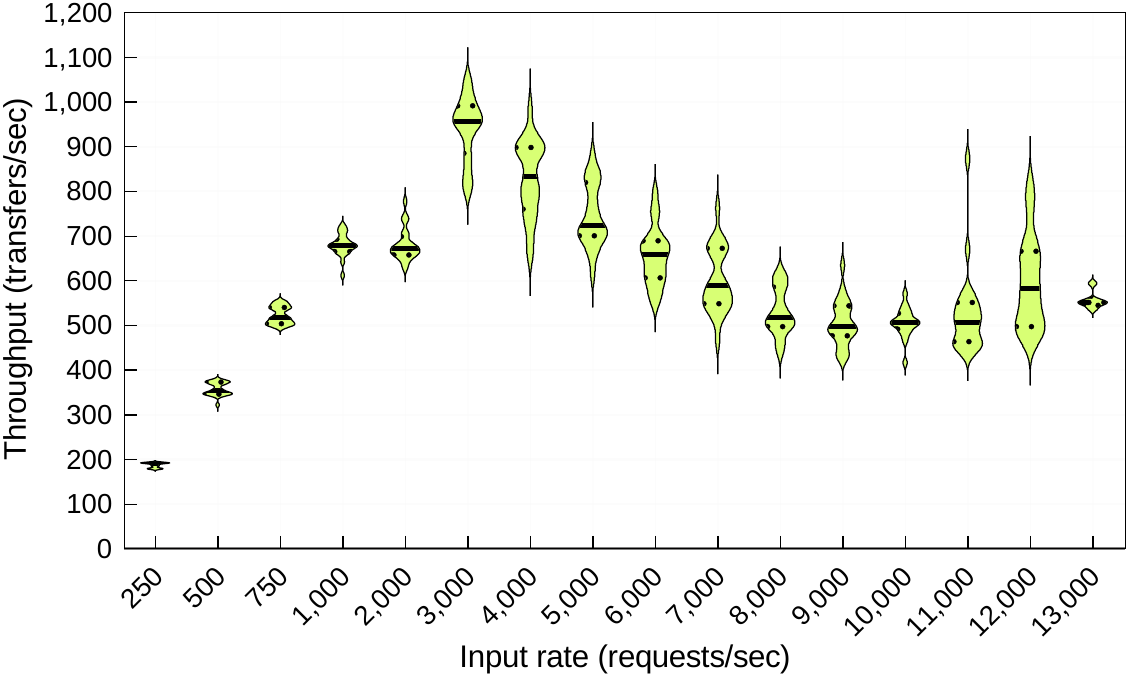}
  \caption{Throughput achieved by the Tendermint blockchain with 5 validator nodes and a network latency of 200ms. Each violin represents the distribution of measurements obtained from 20 executions.}
  \label{fig:tendermint_throughput}
\end{figure}

\textit{\textbf{Tendermint blockchain.}}
We evaluate the throughput of the Tendermint blockchain by using the Hermes Relayer to submit cross-chain transfers (\textit{MsgTransfer}) at input rates ranging from 250 to 14,000 requests per second (RPS) for 15 consecutive blocks. In this setup we test only the blockchain's capacity to initiate cross-chain operations by committing transfer messages. 
It is worth noting that at 14,000 RPS more than 90\% of the requests to initiate a cross-chain transfer fail to be submitted to the blockchain, as we show in Table \ref{table:throughput_errors}.

Figure \ref{fig:tendermint_throughput} shows the throughput achieved by the blockchain with input rates ranging from 250 to 13,000 RPS. We omit results for 14,000 RPS as none of those experiments generated more than 5 consecutive blocks, leading to highly inconsistent results, e.g., throughputs of 10,373 TFPS, 605 TFPS and 0 TFPS. Violins in the plot represent the distribution of the measurements obtained from 20 executions using the same input rate. Inside each violin, the continuous line represents the median throughput and the dotted lines above and below it represent the upper and lower quartiles respectively. We begin by measuring the throughput achieved with 250 RPS, which yields roughly 200 TFPS. Throughput rises as we increase the input rate from 500 RPS (350 TFPS) to 3,000 RPS (961 TFPS), peaking at 961 transfer messages included in the blockchain per second. From 1,000 RPS to 2,000 RPS we observe a slight drop in throughput due to empty blocks, which appear in all experiments with input rates above 2,000 RPS. This is caused by an increase in the amount of processing required by the relayer as the number of requests to be submitted to the blockchain grows. Increased processing time causes the relayer to miss the transaction submission window, leading consensus validators to agree on an empty block.

\begin{figure}[!htb]
  \centering
  \includegraphics[scale=0.76]{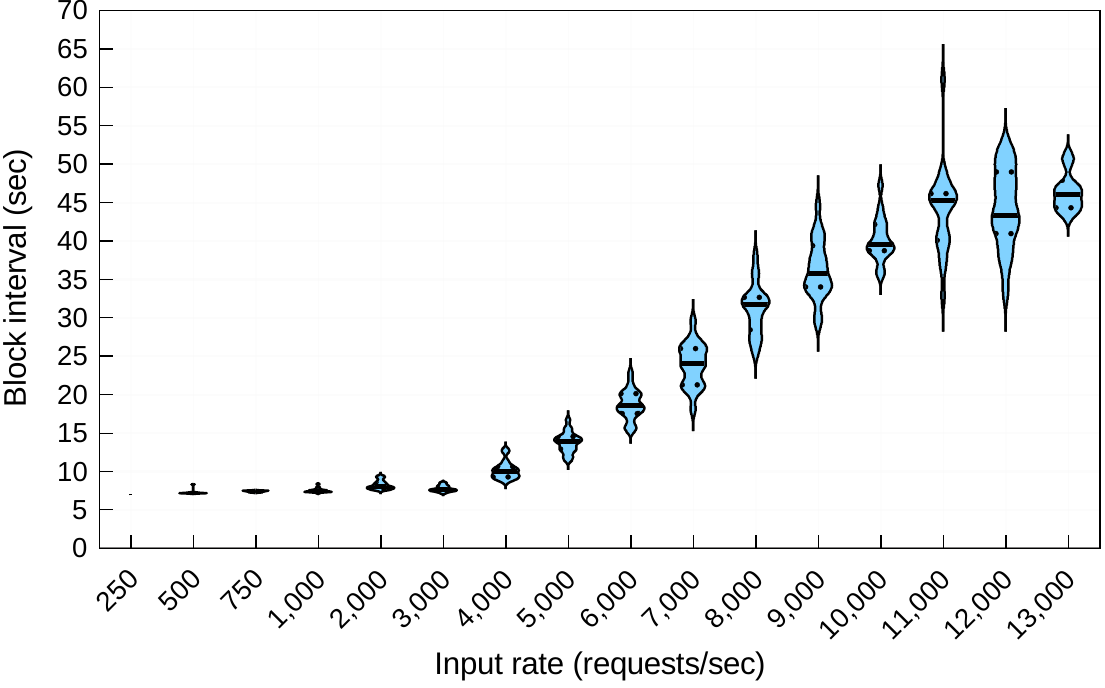}
  \caption{Average time interval between two consecutive blocks according to cross-chain transfer input rates ranging from 250 to 13,000 RPS.}
  \label{fig:tendermint_block_interval}
\end{figure}

For 3,000 RPS and beyond, the variance in measurements is more than twice that of smaller input rates. We also noticed an increase in block intervals as we continue increasing the rate of requests. This increase is shown in Figure \ref{fig:tendermint_block_interval}. At this point, throughput becomes less consistent across each execution and decreases from 830 TFPS at 4,000 RPS to 499 TFPS at 9,000 RPS. From 10,000 RPS to 13,000 RPS measures become even less consistent due to the high amount of errors raised by the relayer application, such as \textit{``account sequence mismatch''} and \textit{``failed tx: no confirmation''}. The \textit{``account sequence mismatch''} errors are caused by the high volume of transactions being submitted. 
High transaction submission rates stresses the blockchain's RPC endpoint, which is also used to query for confirmed transactions. As a consequence, the relayer cannot reliably confirm submitted transactions and increment sequence numbers, leading transaction submission to fail.

We present the rate of failed transfers for each input rate in Table \ref{table:throughput_errors}. From 250 to 10,000 RPS the relayer is able to submit the majority of our requests (99\%) to the blockchain. At 10,000 RPS, 80.17\% of our cross-chain transfer requests reach the blockchain's transaction pool. From those, 98.3\% are included in the chain. Higher request rates greatly decrease the number of transfers submitted and confirmed. At 14,000 RPS, only 8.5\% of our requests make it through to the blockchain, and only 29.2\% of those are committed by the validators.

\begin{table}[!htb]
\renewcommand{\arraystretch}{1.3}
\setlength\tabcolsep{0.5em}
\resizebox{.99\columnwidth}{!}{
\begin{tabular}{cccc}
\textit{\begin{tabular}[c]{@{}c@{}}Input rate \\[-0.4ex] (requests/sec) \end{tabular}} & 
\textit{\begin{tabular}[c]{@{}c@{}}Requests made \\[-0.4ex] to Hermes \end{tabular}} & 
\textit{\begin{tabular}[c]{@{}c@{}}Submitted to\\[-0.4ex] blockchain \end{tabular}} & 
\textit{\begin{tabular}[c]{@{}c@{}}Committed \\[-0.4ex] (from submitted)\end{tabular}} \\ \hline
250 to 9,000 & 18,750 to 675,000 & \textgreater 99\% & \textgreater 99\% \\ \hline
10,000 & 750,000 & 601,300 (80.17\%)  & 591,450 (98.3\%) \\ \hline
11,000 & 825,000 & 319,152 (38.6\%) & 292,424 (91.6\%) \\ \hline
12,000 & 900,000 & 160,343 (17.8\%) & 119,733 (74.6\%) \\ \hline
13,000 & 975,000 & 100,688 (10.3\%) & 51,436 (51\%) \\ \hline
14,000 & 1,050,000 & 90,000 (8.5\%) & 26,360 (29.2\%) \\ \hline
\end{tabular}}
\caption{Execution summary for Tendermint throughput experiments.}
\label{table:throughput_errors}
\end{table}

\textit{\textbf{Hermes Relayer.}} We measure cross-chain throughput by observing the number of cross-chain transfers that Hermes can send from the source to the destination blockchain (\textit{transfer, receive and acknowledge}) within 50 consecutive blocks. We use input rates ranging from 20 RPS to 300 RPS and execute experiments using both one and two instances of the relayer to evaluate its ability to scale. In addition to executing experiments with 200ms network latency, we also performed measurements in a local area network with negligible latency ($<$0.5ms) to demonstrate how it impacts performance. 

The 100 messages used in our transactions consume an average of 3,669,161 gas for transfers, 7,238,699 gas for receives and 3,107,462 gas for acknowledgements. Those requirements vary by at most 1\%, 4.1\% and 7.6\% for transfer, receive and acknowledgement transactions across all of our experiments, respectively. In the Hermes Relayer's configuration, we set the gas price to 0.01 tokens per unit of gas.

\begin{figure}[!htb]
\centering
\includegraphics[width=.99\linewidth]{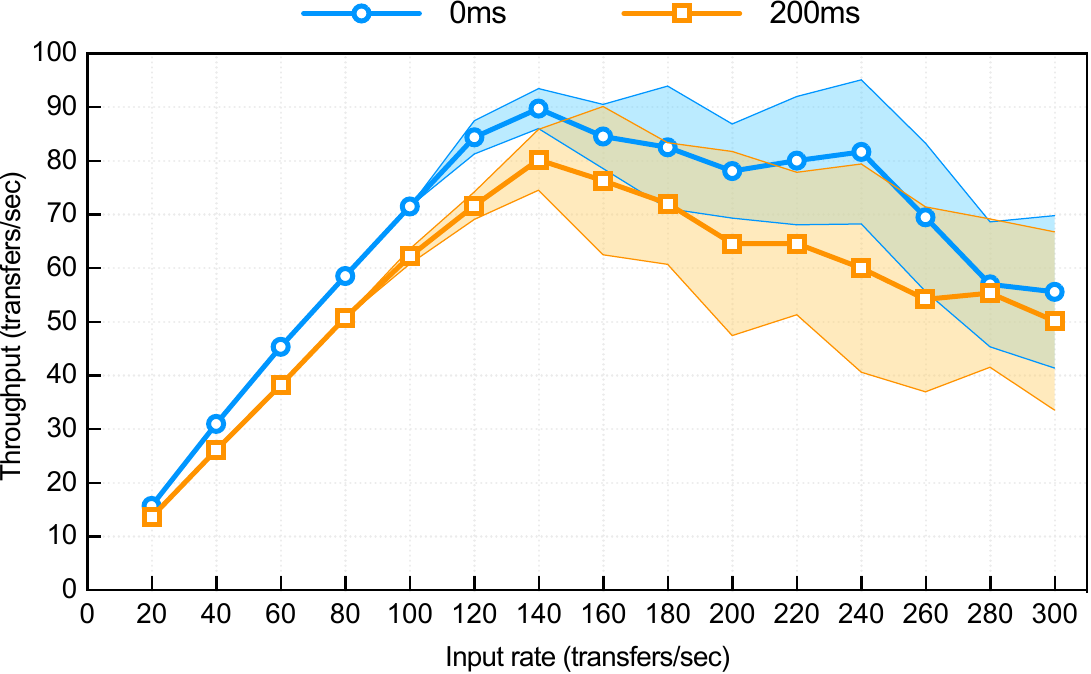}
\caption{Cross-chain transfer throughput achieved by running one instance of the Hermes Relayer. Data points represent the average measurement obtained through 20 executions.}
\label{fig:throughput_1relayer}
\end{figure}

Our results are shown in Figure \ref{fig:throughput_1relayer}. 
Circles indicate measurements with 0ms and squares with 200ms. The error bands surrounding the data points represent the standard deviation of the measurements. Each point represents the average of measurements obtained from 20 executions using the same input rate. With 200ms latency, throughput is consistent across executions from 20 RPS (14 TFPS) to 120 RPS (72 TFPS). Relayer performance peaks at 140 RPS, being able to complete 90 cross-chain transfers per second, and drops with further increase in input rate. 

Our findings show a consistency between the rate of requests and throughput for input rates 20, 40, 60, 80 and 100 TPS, with little variance between the executions within the same scenario. However, as we increase the input rate and approach the maximum throughput of the relayer application in our experiments, the difference between measurements increases. Peak throughput is observed with an input rate of 140 TPS, reaching 90 TFPS with 0ms and 80 TFPS with 200ms latency, a difference of 11\%. Above 140 TPS, throughput gradually decreases as we increase input rate, dropping to 50 TFPS (200ms) and 56 TFPS (0ms) at 300 RPS.

\begin{figure}[!htb]
  \centering
  \includegraphics[width=.99\linewidth]{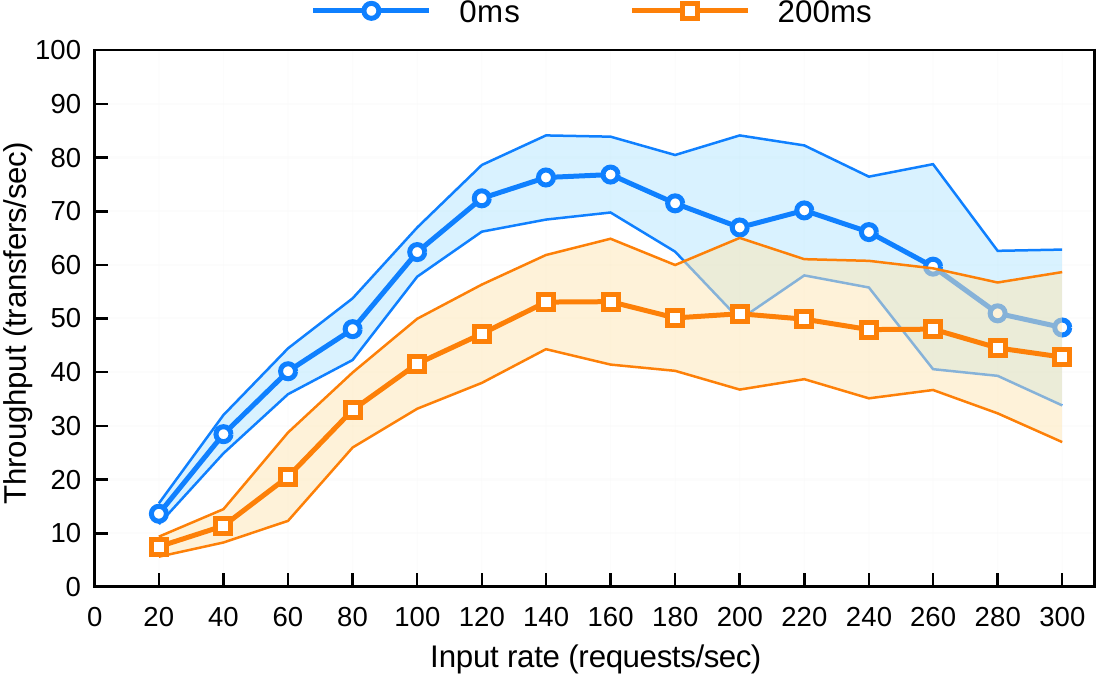}
  \caption{Cross-chain transfer throughput achieved by running two relayers for a single channel. Data points represent the average measurement obtained through 20 executions.}
  \label{fig:throughput_2relayer}
\end{figure}

\begin{figure}[!htb]
  \centering
  \includegraphics[width=0.99\linewidth]{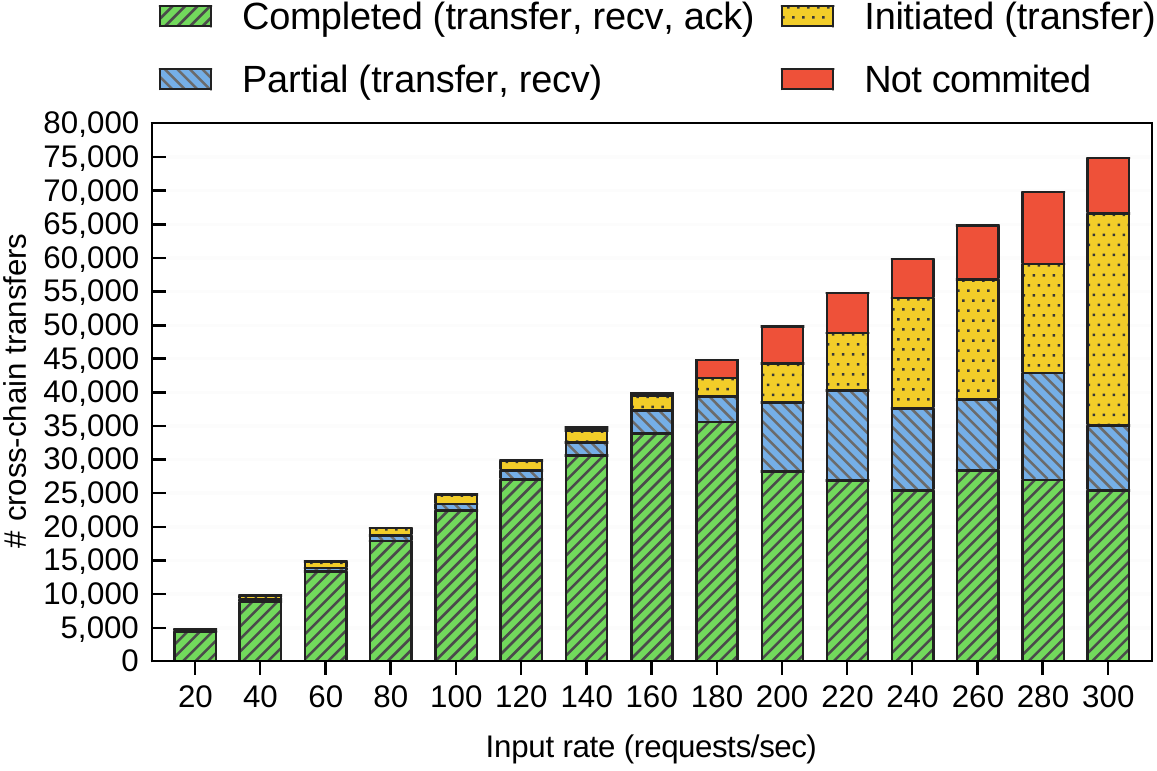}
  \caption{Average number of messages included in the source and destination blockchains within a 50 blocks interval using one relayer.}
  \label{fig:messages_1relayer}
\end{figure}

\begin{figure}[!htb]
  \centering
  \includegraphics[width=.99\linewidth]{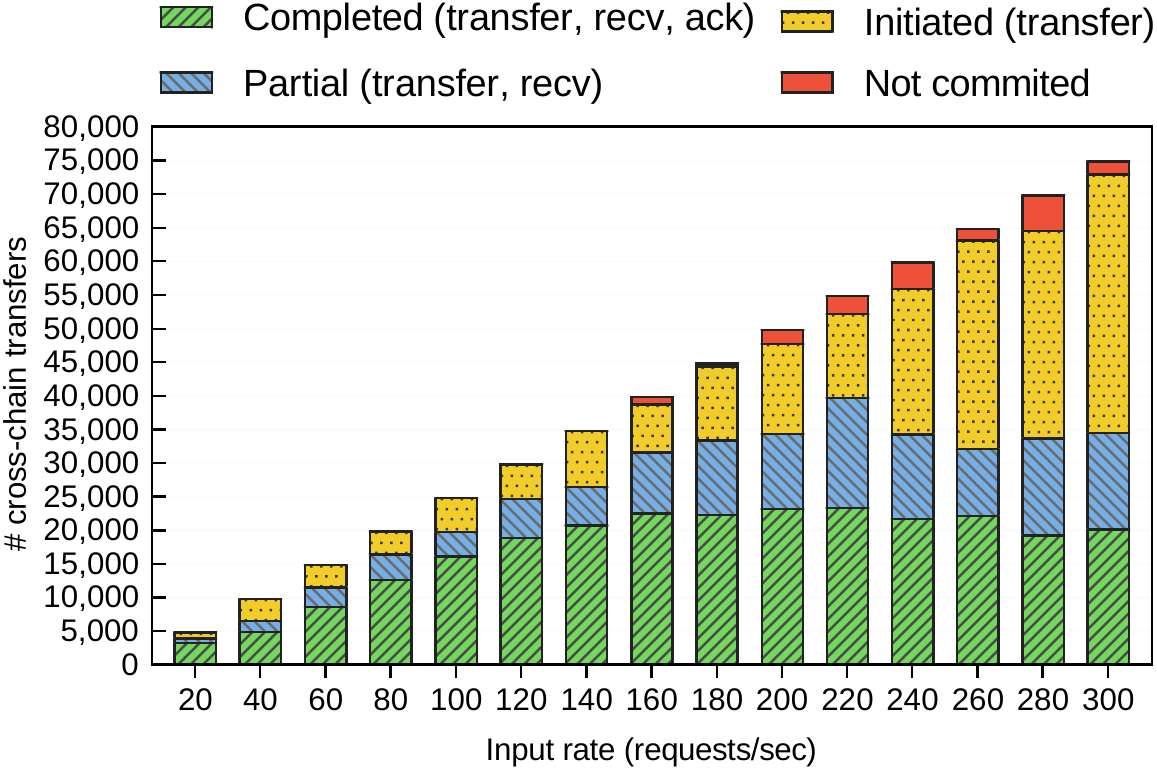}
  \caption{Average number of messages included in the source and destination blockchains within a 50 blocks interval using two relayers.}
  \label{fig:messages_2relayer}
\end{figure}

We also repeated our experiments using two instances of Hermes instead of one. Our goal was to measure the system's throughput with multiple relayers working for a single cross-chain channel. Surprisingly, we got a maximum throughput of just 77 TFPS with 200ms and 53 TFPS with 0ms at an input rate of 160 RPS. 
Comparing these results with our setup using a single relayer, we got a peak performance 14\% and 33\% lower for network latencies of 0ms and 200ms respectively. Upon investigating the logs generated by Hermes, we observed several \textit{``packet messages are redundant''} errors, e.g., 23,020 times for 100 RPS. This happens when both relayers attempt to deliver the same messages to the blockchain, causing the first to succeed and the second to fail. This issue happens due to design choices of the relayer application, such as the inability of relayers to communicate and coordinate packet delivery. The Interchain Standard that defines Relayer Algorithms (ICS 18)~\cite{ics18} does not discuss scalability, a decision which can be detrimental to engineering and implementation. For instance, the standard neither considers nor discusses the use of multiple cross-chain channels for scalability. It also lacks a discussion on how to scale relayers within a single channel. We believe that basic scaling specifications would benefit the protocol if included in the standards. In short, this problem impairs throughput in setups with multiple relayers. Additionally, it also leads relayer operators to pay fees in an attempt to deliver redundant packets.

An alternative to increase cross-chain throughput would be to establish separate cross-chain channels for each relayer to relay on, however, there are downsides to this approach. Tokens sent from a source to a destination blockchain through different channels are represented in the blockchain using different denominations and are not fungible. Additionally, while multiple uncoordinated relayers have a negative impact on throughput, they may maintain network liveness if a subset of them stops working.

Besides throughput, we have collected data regarding transfer completion status during experiments with 200ms latency. We measured the number of transfers that were completed (\textit{transfer}, \textit{receive}, \textit{acknowledge}), partially completed (\textit{transfer},\textit{receive}), only initiated (\textit{transfer}) and those that were not committed to the source blockchain. We present the average message completion status for experiments with one relayer in Figure \ref{fig:messages_1relayer} and two relayers in Figure \ref{fig:messages_2relayer}.

At the lowest input rate of 20 RPS we have submitted 5,000 transfers over a time frame of 50 blocks. At 300 RPS we have submitted 75,000 transfers over the same time frame. Up to 160 RPS, the majority of our transfer requests ($>$99.9\%) were committed to the blockchain with both one and two relayers. With two relayers, however, a portion of the committed transfers are initiated, but fail to be delivered before the end of the experiment due to lower throughput caused by the previously discussed message redundancy errors. For 180 RPS and above, the percentage of completed transfers decreases with both one and two relayers. As we show in our latency analysis later in this section, messages contained within a block are processed sequentially, leading to longer confirmation times as we increase the number of transfers in a block. This leads to two observations in experiments with inputs rates of 180 RPS and above. First, transfers start getting completed, i.e, having the acknowledgement message included in the source chain, only several blocks after the experiments have been initiated. Second, transfers that are submitted several blocks after the start of the experiment were unable to be completed before the blockchain generated the 50th block and ended up either partially completed or only initiated. It is worth emphasizing that we measure the performance achieved by the relayer within a bounded time frame. If given sufficiently large timeouts and enough time after submission all valid transfers can be completed.

\begin{figure*}[!htb]
  \centering
  \includegraphics[width=0.70\textwidth]{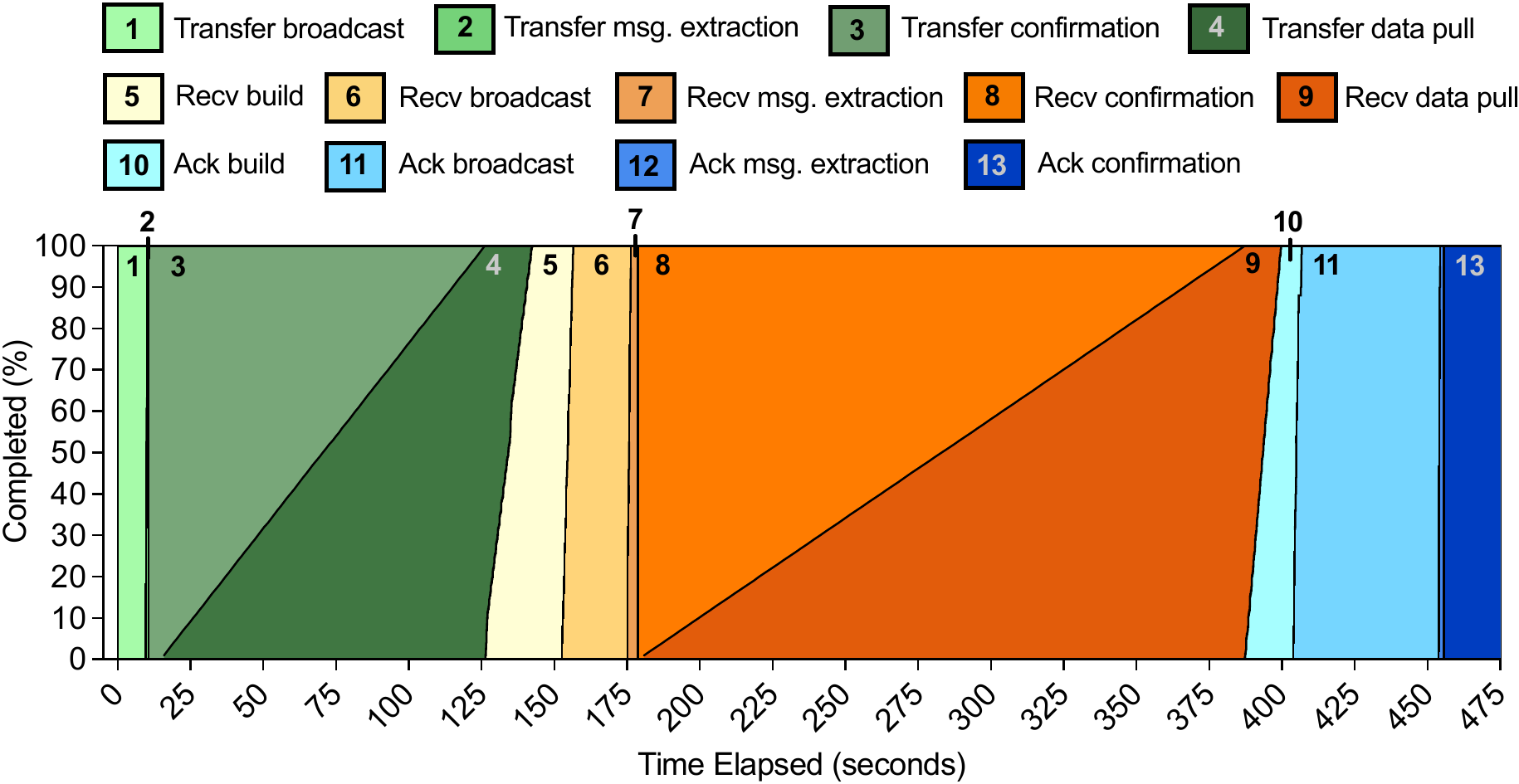}
  \caption{Breakdown of the operations executed to process of 5,000 cross-chain transfers submitted within 1 block.}
  \label{fig:completion_latency}
\end{figure*}

\subsection{Latency}

To measure the latency of cross-chain communication, we break down cross-chain transfers into 13 steps, as shown in Figure~\ref{fig:completion_latency}. A transfer operation starts with broadcast of the \textit{transfer} message and ends with the confirmation of the \textit{acknowledgement} message. Our goal here is to observe the time required by the Hermes Relayer to complete each step of a cross-chain transfer operation. 

We submitted 5,000 cross-chain transfer requests within one block and analyzed the logs generated by Hermes. We show a detailed breakdown of the execution of the 5,000 transfers in Figure \ref{fig:completion_latency}. We divide each of the three steps of a complete successful cross-chain transfer (\textit{transfer, receive, acknowledge}) further into \textit{build}, \textit{broadcast}, \textit{message extraction}, \textit{message confirmation} and \textit{data pull}. Steps are numbered according to the order in which they are executed by the relayer. 

A step starts being processed when the area marked by its number first appears in the figure. As operations get completed, the area moves towards the top. When a step succeeds another, the area of the step that is currently being processed is drawn on top of the previous one. For example, step 4 (\textit{Transfer data pull}) starts being processed at 16 seconds, is 50\% completed (2,500 out of 5,000) at 75 seconds and finishes when 126 seconds have elapsed. Steps such as \textit{message extraction} and \textit{message confirmation} are completed near instantly for all the requested transfers, leading to a vertical line in the figure. Others, such as \textit{build} require more processing and have their completion percentage increase gradually.

We start measuring the latency of operations from the moment \textit{transfer} messages are broadcast. After a message is broadcast, the relayer extracts status information from the blockchain (\textit{message extraction}), confirms that it has been committed (\textit{confirmation}) and requests the data contained in it to the blockchain (\textit{data pull}) to build the next message required by the protocol.
Overall, steps required to process the \textit{transfer} message consume 126 seconds or 27.6\% of the total processing time, steps related to the \textit{receive} message require 261 seconds (57.3\%) and steps related to \textit{acknowledgement} message are processed in 68 seconds (14.9\%). The first transfer operation to be completed out of the 5,000 we submitted required 455 seconds from \textit{transfer broadcast} to\textit{ acknowledgement confirmation}.

Blocks are handled sequentially by the relayer, leading it to process the first step for every \textit{transfer} message within the block before moving on to the next.
We observe that the main bottleneck in the process of cross-chain communication is caused by the operations that retrieve data from the \textit{transfer} and \textit{receive} messages included in the blockchain, namely \textit{transfer data pull} and \textit{recv data pull}. The former corresponds to 24\% (110 seconds) of the time spent processing the 5,000 cross-chain transfers. The latter corresponds to 45\% (207 seconds). Together, those operations require 317 seconds to be completed, which is roughly 69\% of the total processing time. For the duration of those operations, Hermes is using RPC to query blockchain nodes for data. The reason why the \textit{transfer data pull} and \textit{recv data pull} operations correspond to over two thirds of the the time required for cross-chain communication lies in Tendermint's RPC implementation. Tendermint is unable to process queries in parallel, requiring the relayer to wait while its requests for data are processed one by one. This leads to completion latencies in the order of several minutes.

\begin{figure}[!htb]
  \centering
  \includegraphics[scale=0.79]{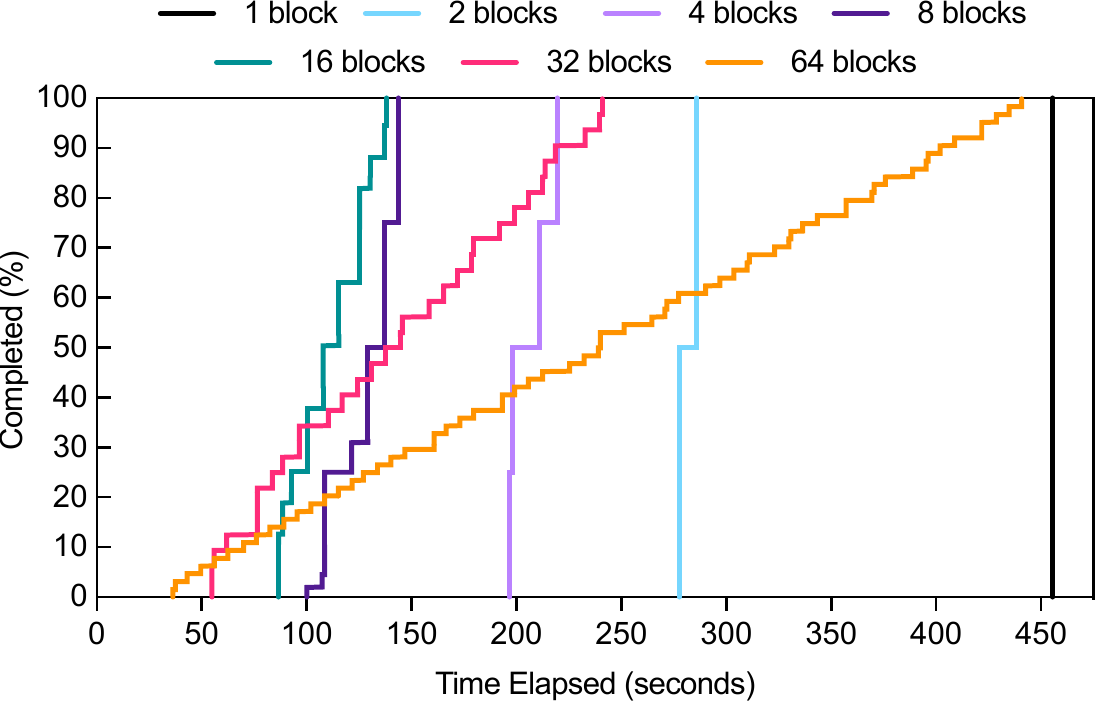}
  \caption{Completion of 5,000 cross-chain transfers over time based on seven distinct transaction submission strategies.}
  \label{fig:completion_times_5k}
\end{figure}

We conducted experiments using different transaction submission strategies in an attempt to reduce the completion latency of cross-chain transfers. For this purpose, we evenly divided the total number of cross-chain transfer requests into different intervals of blocks, ranging from 1 to 64.
Figure \ref{fig:completion_times_5k} shows how different submission strategies affect completion latency. Submitting 5,000 transfers within 1 block, as also shown in detail in Figure \ref{fig:completion_latency}, leads to a 455 seconds completion latency for all 5,000 transfers. Dividing transfer submission into 2 blocks yields a 286 seconds completion latency, a 37\% reduction. As we double the interval over which we submit transfers, completion latency is further reduced to 219 seconds (4 blocks), 143 seconds (8 blocks) and 138 seconds (16 blocks). We observe that spreading submissions further beyond this point increases completion latency, rather than reducing it further, to 240 seconds (32 blocks) and 441 seconds (64 blocks). 
We conclude that submitting cross-chain transfers in large batches, while easier to carry out, as one does not have to coordinate submissions over multiple blocks, severely increases completion latency. Increasing the submission window from 1 to 16 blocks provides a 70\% reduction in completion latency. We note however, that spreading transfer submission over larger block intervals has an inverse effect, increasing completion latency from 138 seconds to 441 seconds, a 320\% increase. Additionally, splitting the submission into multiple blocks causes transfers to start being completed sooner, in increments proportional to the number of transfer requests inside each block.

\section{Deployment Challenges}\label{sec:challenges}

In this section we summarize technical challenges faced throughout this work with two goals in mind. First, to further give grounds for some of our design choices and elaborate on the impact they had on our benchmarks. Second, to bring attention to those issues and contribute to the improvement of the affected systems.

\begin{itemize}
    \item \textit{Timestamp mismatch}: Analysis of experimental data revealed a difference between the timestamp of events registered by the Tendermint blockchain and those registered by the relayer application. We observed that events logged by the blockchain are a few seconds behind, e.g, the timestamp for a transaction broadcast recorded by the relayer is set after the same transaction has already been included in the blockchain according to the block timestamp.
    This mismatch in the timestamps recorded by both applications may lead to incorrect metrics if both sources of information are utilized together. For this reason we chose to rely on the timestamps recorded by Hermes for our cross-chain performance analysis.

    \item \textit{Account sequence mismatch}:
    When attempting to create transactions that contain cross-chain messages through the relayer command line interface (CLI), we were unable to submit more than one transaction per block for each user account (public key). Attempting to submit a new transaction from the same account while a previously submitted transaction has not yet been confirmed yields the \textit{``Account sequence mismatch''} error.
    This can be circumvented by issuing multiple transfer messages in the same transaction, allowing a user to perform multiple payments without having to wait for each transaction to be confirmed. In practice, however, this workaround has the downside of requiring the payer to accumulate outgoing transfers in order to include them all in the same transaction. Furthermore, users may be still unable to issue new transactions if a previous one remains awaiting confirmation. To overcome this limitation we chose to use multiple user accounts to submit transactions, allowing us to control the volume of messages submitted per block by changing the number of concurrent users issuing transactions containing transfer messages.
    
    \item \textit{Transaction data collection}:
    During the development of our tool we explored the options provided by the available data endpoints in order to choose the queries that best suited our needs. 
    Retrieving the list of the transaction hashes associated with the transactions included in a block allows us to query the blockchain for those specific transactions and their cross-chain message contents.
    Given the large amount of transactions we generate (e.g., 2250) during benchmark execution, one of our main goals was to minimize the number of queries required for data collection to reduce execution time.
    Upon a careful examination of the queries supported by the Tendermint RPC and the Gaia blockchain we were unable to find an option that returned the list of transaction hashes in a specific block without returning a substantial amount of additional transaction  data. Both the {\tt tx\_search} query parameter from Tendermint RPC and the {\tt query txs --events tx.height=X} from the Gaiad CLI return a large amount of information. For instance, querying a local node for the content of a block that includes 20 transactions with 100 \textit{MsgTransfer} each returned 331,706 lines of output and took an average time of 2.9 seconds to complete using our setup. Querying a block with 20 transactions containing 100 \textit{MsgRecvPacket} each returned 579,919 lines of output and took an average time of 5.7 seconds. 
    It is essential for us to collect the hashes of the transactions contained within a block to perform the our analysis, therefore we chose to use the aforementioned queries despite their negative impact on performance. As such queries may return a substantial amount of data, we had to deal with pagination as some blocks can be too large to fit in a single request. 

    \item \textit{WebSocket space limit}:
    When blocks are minted, the relayer application queries the blockchain to retrieve their contents and identify transactions containing pending IBC messages. If the amount of data to retrieve exceeds the Tendermint Websocket maximum message size (16MB), the relayer emits the \textit{``Failed to collect events''} error. 
    This issue happens every time a block contains a large amount of IBC events, such as transfers, that need to be processed. We observed that when this error is raised, if the relayer packet clear interval is set to 0, pending IBC transfers neither get completed nor fail even when their timeout is exceeded. 
    To analyze the impact this has on packet relaying, we ran an experiment and adjusted the parameters with the intent to trigger this error. 
    We generated a block containing 1,000 cross-chain transactions with 100 IBC transfers each, causing the \textit{``Failed to collect events''} error. From the transactions in this block, 2.5\% were completed successfully, 15.7\% timed out and 81.8\% got stuck, meaning they were committed on the source chain but neither were relayed successfully or timed out, even after four times the number of blocks required for timeout were appended to the chain.
    Shortly after, we submitted multiple transactions containing a single IBC transfer. All single message transfers submitted after the error were committed to the source blockchain, but were not delivered by the relayer, causing them to time out. This indicates that the WebSocket error not only prevents transactions that failed to be collected from being completed, but also impacts future transactions.

    \item \textit{Incomplete logging for blockchain data retrieval}:
    
    By analyzing the execution logs generated by the Hermes Relayer we were able to collect data regarding the operations performed by the application and their timestamps. However, only a fraction of the operations that query the blockchain RPC endpoints for IBC message data are recorded in the logs. This is the operation we are most interested in, given that it is the main bottleneck in the process of cross-chain communication. We observed that the retrieval of data from the first block of transactions in our experiments is correctly logged, but not that of transactions included in subsequent blocks. Upon verification we confirmed that transactions included in subsequent blocks are queried for their data, as their packets get built and delivered by the relayer, despite the data pull operation not being recorded in the logs. We suspect this is caused by an implementation issue.

\end{itemize}
\section{Related Work}

Most previous works on empirical blockchain performance analysis have essentially targeted the execution of workloads in isolated blockchains. Those workloads are usually designed to stress test specific blockchain functionalities such as the consensus protocol, smart contract language or a decentralized application. Hyperledger Fabric has been extensively analyzed using micro-benchmarks~\cite{baliga2018fabric}, custom workloads~\cite{thakkar2018performance, nakaike2020hyperledger}, variable transaction submission rates~\cite{kuzlu2019performance} and blockchain performance benchmarking tools~\cite{wang2021xbcbench, dinh2017blockbench, saingre2020bctmark}. Following, Ethereum has also been the target of several studies analyzing the platform's throughput and latency~\cite{dinh2017blockbench, wang2021xbcbench, sedlmeir2021dlps}. Both Hyperledger Fabric and Ethereum have also undergone fault tolerance and on-chain smart contract performance analysis~\cite{saingre2020bctmark, aldweesh2019opbench, dinh2017blockbench}. 

Interoperability in the domain of computer systems is an established topic in the literature~\cite{viho2001towards, leal2019interoperability}, however, works that discuss interoperability, performance and conformity of cross-chain communication solutions only recently began to appear ~\cite{belchior2023hephaestus,belchior2022interoperabilitysolution,koens2019assessing}. 
For instance, a framework has been proposed to systematically assess the interoperability degree of cross-chain communication solutions, including cross-chain performance as a guideline~\cite{belchior2022interoperabilitysolution}.
Similarly, another work proposes a set of properties to evaluate interoperability solutions for distributed ledgers, however, the only performance metric considered is scalability~\cite{koens2019assessing}. 
Lastly, Hephaestus~\cite{belchior2023hephaestus} generates models to identify malicious behavior in cross-chain communication. 
However, unlike ours, none of those previous works provides a concrete framework to guide the process of empirically evaluating the performance of cross-chain communication solutions.
In addition, we provide the first comprehensive evaluation of cross-chain communication using the IBC protocol.

\section{Conclusion}

This work presented a novel framework to guide empirical performance evaluation of cross-chain communication protocols. We implemented our framework as a tool to conduct the first comprehensive analysis of throughput, latency and relayer scalability of cross-chain communication using the IBC protocol and the Hermes Relayer.

Through our findings we highlight the importance of studying the performance of such systems. Aided by our framework we found several issues that hinder the performance of components related to cross-chain communication such as the Cross-chain Communicator and the Cross-chain Data Connector. For instance, using two relayers to relay for a single IBC cross-chain channel decreases throughput by 33\%. In addition, the lack of support for parallel query processing in the blockchain's RPC endpoint causes data retrieval to consume 70\% of the time required to complete 5,000 cross-chain transfers. Those findings bring attention to the impact that those components have on cross-chain performance and present opportunities for the improvement of cross-chain communication protocols.

Finally, to assist future research and development, we make available a 158GB dataset of execution logs obtained through our experiments. 
\section{Acknowledgement}
This work was partially supported by the Australian Research Council (ARC) under project DE210100019.

\bibliographystyle{IEEEtran}
\bibliography{references}
\end{document}